# Correction of Residual Errors in Configuration Interaction Electronic Structure Calculations

Jerry L. Whitten

Department of Chemistry

North Carolina State University

Raleigh, NC 27695 USA

email: whitten@ncsu.edu

**Abstract**

Methods for correcting residual energy errors of configuration interaction (CI) calculations of molecules and other electronic systems are discussed based on the assumption that the energy defect can be mapped onto atomic regions. The methods do not consider the detailed nature of excitations, but instead define a defect energy per electron that that is unique to a specific atom. Defect energy contributions are determined from calculations on diatomic and hydride molecules and then applied to other systems. Calculated energies are compared with experimental thermodynamic and spectroscopic data for a set of forty-one mainly organic molecules representing a wide range of bonding environments. The most stringent test is based on a severely truncated virtual space in which higher spherical harmonic basis functions are removed. The errors of the initial CI calculations are large, but in each case, including defect corrections brings calculated CI energies into agreement with experimental values. The method is also applied to a NIST compilation of coupled-cluster calculations that employ a larger basis set and no truncation of the virtual space. The corrections show excellent consistency with total energies in very good agreement with experimental values. An extension of the method is applied to $d^m s^n$ states of Sc, Ti, V, Mn, Cr, Fe, Co, Ni and Cu, significantly improving the agreement of calculated transition energies with spectroscopic values.

**I. Introduction**

The accuracy of electronic structure calculations by configuration interaction (CI) depends on the completeness of the single particle basis and the completeness of the set of configurations used in the expansion of the wavefunction. As systems increase in size it becomes increasingly difficult to obtain energies close to the exact values. There is a vast literature on CI approaches ranging from perturbation methods that generate configurations and evaluate energies efficiently to methods for partitioning large systems into localized electronic subspaces or ways to balance errors in systems that are being compared.[1-39] Relatively few configurations are required to dissociate molecules correctly or to create proper spin states, but dynamical correlation effects, particularly those associated with angular correlation, require higher spherical harmonic basis functions and this leads to a rapid increase in number of interacting configurations. Finding more efficient ways to treat large systems by coupled-cluster[22-24] and multireference methods[9-11,25-27] and methods that use non-orthogonal molecular orbitals[28-29] are continuing research areas in the quest to find increasing accurate descriptions of ground and excited states of molecules and materials.

In this work, we assume a CI calculation has been carried out on an electronic system. The calculation is assumed to be sufficiently accurate to describe spin states and to capture important static correlation contributions but may be deficient in its single-particle basis or completeness of the CI. The objective is to correct the energy error to agree more closely with the exact energy of the system determined from experimental thermodynamic and spectroscopic data.

The paper is organized in several parts: (1) The ideas underlying the correction methods are discussed; (2) Two methods of error correction are applied to a data set of forty-one mainly organic molecules containing C, N, O, F and H in a variety of bonding environments; (3) One of the error correction methods is applied to a published NIST data base of coupled-cluster calculations[2]; and (4) Error corrections are applied to ground and excited states of transition metal atoms Sc, Ti, V, Mn, Cr, Fe, Co, Ni and Cu corresponding to $s^2d^n$, $s^1d^{n+1}$, $s^0d^{n+2}$ occupancy. The transition metal states have substantially different correlation energies, and the objective is to bring calculated transition energies into agreement with spectroscopic values.

Some of the same ideas have been discussed in a previous paper (JCP 2020)[5] where the objective was to correct lower-level CI calculations to match more closely the energies of higher-level calculations. The present work extends the methods in new directions and provides a more stringent test of the accuracy.

## II. Methods

For a given electronic system, we begin with a CI calculation that is sufficiently accuracy to account for major static correlation contributions. As noted, errors in the calculation may exist from incompleteness of the basis set or incomplete configuration interaction. For example, missing from the expansion may be configurations containing higher spherical harmonic components required for angular correlation. The first requirement of a correction method is to account for missing local contributions to the correlation energy.

The correction is formulated as an energy defect per electron that depends on the location in space. For a system of N electrons, the energy defect is

$$E^{corr} = \int \gamma(x,y,z) \rho(x,y,z) dv$$

where $\rho(x,y,z)$ is the electron density and $\gamma$ is an energy defect that varies within the molecule. The Hohenberg-Kohn theorem would allow $\gamma$ to be expressed as a functional of the density,[40] however, we wish to develop an argument in a much simpler direction. We consider below two different arguments that lead to the same general conclusion. Since the objective is to match the exact total energy, $\gamma$ factors are intended to account for the total energy defect of the CI treatment including missing electron correlation, single-particle basis set deficiencies and relativistic effects. We shall refer to $\gamma$ subsequently as an energy defect factor.

### Argument 1

Density expansions can be determined by minimizing a rigorous 2-particle error bound,[41]

$$< \rho(1) - \rho'(1) | r_{12}^{-1} | \rho(2) - \rho'(2) > \geq 0$$

where $\rho$ is a single-determinant (SCF) density defined by occupied molecular orbitals, $\varphi_p$, and basis functions, $f_i$,

$$\rho = \sum_p \varphi_p \varphi_p = \sum_{i,j} w_{ij} f_i f_j$$

and $\rho'$ is a proposed approximation

$$\rho' = \sum_{i,j}^{i,j \in M} \lambda_{ij} f_i f_j \qquad (i, j \text{ on same atomic site M})$$

The coefficients $\lambda_{ij}$ are chosen to minimize the error bound and then renormalized to give the exact number of electrons. Limiting the expansion to basis functions on the same nucleus is a major restriction, but as discussed previously such expansions can accurately approximate the total Coulomb interaction energy.[5,41-46] Table I illustrates the quality of the expansion for several molecules for the basis sets used in the present work. Using the expansion gives the correction

$$E^{corr} = \sum_{i,j}^{i,j \in M} \lambda_{ij} \int \gamma f_i f_j dv = \sum_{i,j}^{i,j \in M} \lambda_{ij} <f_i|\gamma|f_j>$$

We now define an average $\gamma_M$ for basis functions on the same nucleus

$$E^{corr} = \sum_{M}^{nuclei} \gamma_M \sum_{i,j}^{i,j \in M} \lambda_{ij} <f_i|f_j> = \sum_{M}^{nuclei} \gamma_M P_M$$

where $P_M$ is the population of electrons on nucleus $M$ defined by the density expansion.

**<u>Argument 2</u>**

Consider the exact expansion

$$\rho = \sum_p \varphi_p \varphi_p = \sum_{i,j} w_{ij} f_i f_j$$

and

$$E^{corr} = \sum_{i,j} w_{ij} \int \gamma f_i f_j dv = \sum_{i,j} w_{ij} <f_i|\gamma|f_j>$$

If $f_i$ and $f_j$ are on different nuclei $M$ and $N$, or on the same nucleus $M=N$, then we define $\gamma = \tfrac{1}{2}(\gamma_M + \gamma_N)$, i.e., the defect correction for the overlap region is taken as the average of the two atomic contributions. This gives

$$E^{corr} = \sum_{M}^{nuclei} \gamma_M \sum_{i,j}^{i,j \in M} w_{ij} <f_i|f_j> = \sum_{M}^{nuclei} \gamma_M P_M$$

where $P_M$ is the population. Because of the definition of the overlap contribution, the final population in the second approach is the same as the Mulliken population. The populations in Approach 1 and Approach 2 will be numerically different, however, except for homonuclear diatomic molecules.

**Table 1.** Expansion of electron densities based on minimization of the rigorous bound $\varepsilon = <\rho(1) - \rho'(1) | r_{12}^{-1} | \rho(2) - \rho'(2)> \geq 0$ where $\rho$ is the exact SCF density and $\rho'$ is an expansion containing only basis functions on the same site (see text). Values in parentheses are for a $\rho'$ expansion using coefficients from a Mulliken approximation for basis function products on different nuclei.

|  | $<\rho(1) | r_{12}^{-1} | \rho(2)>$ | $<\rho'(1) | r_{12}^{-1} | \rho'(2)>$ | $\varepsilon$ | % error[a] |
|---|---|---|---|---|
| Ethylene | 70.3814 | 70.3726 (70.3471) | 0.0088 | 0.0125 |
| Acetylene | 60.6961 | 60.6948 (60.6505) | 0.0013 | 0.0021 |
| Benzene | 312.2793 | 312.2612 (312.1938) | 0.0181 | 0.0058 |
| Furan | 269.8652 | 269.8451 (269.7715) | 0.0201 | 0.0074 |
| $C_6H_5$-$NH_2$ | 407.5608 | 407.5395 (407.4531) | 0.0213 | 0.0052 |
| Glycine | 315.7287 | 315.7069 (315.6273) | 0.0218 | 0.0069 |
| Glyoxal | 212.0970 | 212.0848 (212.0314) | 0.0122 | 0.0057 |
| $H_2O$ | 46.7273 | 46.7242 (46.6978) | 0.0031 | 0.0067 |
| FHCO | 172.6083 | 172.5957 (172.5507) | 0.0126 | 0.0073 |

[a] The % error $= 100\varepsilon / <\rho(1) | r_{12}^{-1} | \rho(2)>$; energies are in hartrees.

The second approach is extraordinarily simple and instead of working with the total density, we can define a modified Hamiltonian for the system as

$$H'' = H + \sum_i h_i''$$

where H is the exact Hamiltonian and $h_i''$ is a one electron operator that carries the defect contribution. The $h_i''$ is defined by its matrix elements

$$<f_i|h''|f_j> = \tfrac{1}{2}<f_i|f_j>(\gamma_M+\gamma_N)$$
$$f_i \; \varepsilon \; M \quad f_j \; \varepsilon \; N$$

The expectation value of $H''$ then includes the correction. The correction factors slightly affect the iterations of an SCF calculation transferring charge to the atom with larger $\gamma$. If there were no difference in values of $\gamma_M$ there would be no change in the Fock operator since for orthogonal orbitals occupied $\varphi_p$ and virtual $\varphi_q$ since $<\varphi_p|\gamma|\varphi_q>=\gamma<\varphi_p|\varphi_q>=0$. The energies reported in subsequent tables contain the self-consistent-field and CI contributions.

### III. Many-electron calculations and virtual space reductions

As basis sets and the configuration interaction method approach completeness, uncertainties in the energy-defect factor corrections decrease. We shall later examine a few higher-level couple-cluster calculations to demonstrate this point. In this section, however, we are interested in a stringent test of the energy-defect method by sharply reducing the size of the virtual space and the CI. Calculated energies are compared with exact experimental values from thermodynamic and spectroscopic data. Basis sets are described in Appendix I. Hartree-Fock quality expansions of 1s, 2s and 2p orbitals for C, N, O, and F are employed plus additional $s', s'', p', p'', d$ functions to allow polarization and correlation contributions; the basis for H is $1s, s', p'$. The multi-reference CI method used for all calculations has been described previously and is summarized in the Appendix.

The virtual space for all molecular calculations is determined as follows:
a) A SCF calculation on the molecule of interest is carried out using the full basis. A virtual space is created by removing the diffuse functions $s''$, $p''$, the $d$ functions, and hydrogen $p$-type functions, thus, leaving only a double-zeta type basis for the virtual space. This can be done either by localization or as in the present work by carrying out a SCF calculation for the virtual space with the unwanted functions removed. The virtual molecular orbitals are orthogonalized to the occupied SCF orbitals and to other virtual orbitals

b) A CI calculation on the system with the set of occupied molecular orbitals and the reduced set of virtual orbitals is carried out to obtain a CI energy, $E^{CI}$. We refer to the difference $E^{error} = E^{CI} - E^{exact}$ as the energy defect to be captured by the defect factors $\gamma_M$.

Proceeding in this way avoids confusing the polarization and correlation roles of the omitted functions since the full-basis and polarization effects are included in the 1-det SCF calculation.

**Determination of $\gamma$**

We now consider the determination of $\gamma$ values for molecules containing C, N, O, F and H. The CI calculations use the truncated virtual space defined in the previous section. For a given molecule, it is always possible to find values for $\gamma$ that correct the calculated CI energy to match the exact energy. One way of proceeding would be to consider a reference set of molecules representing different bonding environments and determine average values $\gamma_M$ for each atomic component $M$. The usefulness of the result would depend on the deviation of individual values from the average. We shall adopt a simpler approach based only on calculations for the diatomic and hydride molecules $M_2$ and $MH_y$ $M = C, N, O, F, H$. The diatomic molecules include sigma and pi type bonding while the hydrides emphasize sigma bonding.

For the homonuclear diatomic molecules, it follows from the equivalence of the nuclei, that $\gamma_M = E^{error}_{M_2} / N_{M_2}$ where $N_{M_2}$ is the total number of electrons in the molecule (including 1s electrons). For the hydride, a new value of $\gamma_M$ is determined assuming the diatomic value for hydrogen, $\gamma_H$. Values for $\gamma_M$ are reported in Table 2 along with an average value for each nucleus $M$. For other molecules, we would expect the optimum values for $\gamma_M$ to lie within or close to the hydride-diatomic limits. We investigate this question in the following section. Improvements in either the basis or many-electron treatment would lead to a smaller hydride-diatomic range and less uncertainty in the energy-defect calculations.

**Table 2**. Energy defect factors, $\gamma_M$, (energies per electron) calculated at the CI level for diatomic, $M_2$, and hydride, $MH_y$, molecules using $< f_i\ |h''|f_j > \ = \ \frac{1}{2} < f_i\ |f_j > (\gamma_M + \gamma_N)$, see text.

| atom | diatomic | hydride | avg |
|------|----------|---------|-----|
| F | 0.035155 | 0.035488 | 0.035321 |
| O | 0.031874 | 0.033971 | 0.032923 |
| N | 0.027339 | 0.031312 | 0.029326 |
| C | 0.025380 | 0.029620 | 0.027500[a] |
| H | 0.010370 | 0.010370 | 0.010370 |

[a] $\gamma_C = 0.027267$ equalizes the error per C in $C_2$ and $CH_4$. This value is used in all subsequent calculations. For other atoms, the differences between the average and equal M errors are negligible.

## IV. Molecular calculations

To evaluate the accuracy of the proposed energy defect corrections, we carry out calculations on forty-one molecules representing different bonding environments. As noted earlier all basis functions are included in the SCF description to provide flexibility and account for polarization contributions. The virtual space is severely truncated, however, as described above; for example, the number of virtual molecular orbitals used in the CI is only 12 for $CH_4$ (reduced from 33) and 45 for benzene (reduced from 117). The resulting CI expansions are therefore relatively small even for single and double excitations from up to 400 reference determinants and a selection threshold of 1x10-7 because of the reduced size of the virtual space. The results are intended as a test of the error correction method when the CI is sufficient to include important static effects, but the total energy of the CI is far from the exact value.

Before considering the full set of molecules, it is useful to focus on a few representative molecules to define the scope of the study. We first consider Approach 2 which makes use of $h''$ to include the defect correction. Calculations and supporting information on which the exact energies are based are reported in Table 3.

**Table 3.** Calculations on representative molecules including details on the exact and calculated energies with and without energy defect factors. Information for other molecules is contained in subsequent tables and the Appendix.

Atoms (experimental)[a]

|  | C | N | O | F | H |
|---|---|---|---|---|---|
| Sum $I_p$ (cm$^{-1}$) | 1030.1085 | 1486.058 | 2043.8428 | 2715.89 |  |
| Energy (hartree) | -37.85577 | -54.61160 | -75.10980 | -99.80707 | -0.50000 |

Molecules (experimental)[a]

|  |  | Atomization E kJ/mol 0 K | ZPE cm$^{-1}$ | Atomization E plus ZPE | Exact E 0 K |
|---|---|---|---|---|---|
| $C_2H_4$ | ethylene | 2225.5 | 10784.7 | 0.89679 | -78.60833 |
| $CH_3F$ | fluoromethane | 1683.5 | 8376 | 0.67938 | -139.84221 |
| $C_6H_6$ | benzene | 5463 | 21392.5 | 2.17822 | -232.31284 |
| $C_4H_4N_2$ | pyrazine | 4488 | 16307.5 | 1.78369 | -264.42425 |
| $C_2O_2H_2$ | glyoxal | 2554.5 | 7868 | 1.00881 | -227.93995 |

Molecules (calculated)

|  |  | (No correction) | | | (Includes correction) | | | % E not recovered |
|---|---|---|---|---|---|---|---|---|
|  |  | SCF | CI | CI - Exact | SCF | CI | CI - Exact |  |
| $C_2H_4$ | ethylene | -78.0582 | -78.2581 | 0.3503 | -78.4167 | -78.6148 | -0.0065 | -1.8 |
| $CH_3F$ | fluoromethane | -139.0966 | -139.3213 | 0.5209 | -139.6027 | -139.8259 | 0.0164 | 3.1 |
| $C_6H_6$ | benzene | -230.7684 | -231.2930 | 1.0199 | -231.7965 | -232.3193 | -0.0065 | -0.6 |
| $C_4H_4N_2$ | pyrazine | -262.7627 | -263.3116 | 1.1126 | -263.8573 | -264.4040 | 0.0202 | 1.8 |
| $C_2O_2H_2$ | glyoxal | -226.6750 | -227.0830 | 0.8570 | -227.5470 | -227.9524 | -0.0124 | -1.5 |

[a] Energies are in hartree units unless specified otherwise.
Experimental values are from NIST, Ref. 2

The error in total energy of the truncated CI calculation (CI- Exact) is initially quite large, but is found to be considerably reduced on inclusion of the error correction $h''$ in the recalculated SCF and CI. The percentage of the energy not recovered is also relatively small. The table shows the difference between uncorrected and corrected SCF energies is very close to the difference between uncorrected and corrected CI energies (to within~ $2 \times 10^{-3}$). Thus, the single-determinant contains nearly the entire correction with only slight additional contributions at the CI level.

Similar results are found for the 41 molecules investigated as shown in Table 4. Although the details are important for completeness, it is helpful to focus on two columns of the table: the large error of the truncated CI calculation (CI- Exact) and error after correction shown in the last column. (CI-Exact). For all molecules, the table shows a very substantial reduction of the error. Figure 1 shows a plot of the energy error recovered by the correction. The average error in the energy not recovered (summing over the absolute value of the individual errors) is 2.1% (97.9% recovery). Atomization energy errors are plotted in Figure 2. The energies are calculated using exact atomic energies; thus, no cancellation of errors is involved. The corrected values have an average error of 1.83%.

As noted earlier, these calculations probe the limit of a large initial CI error due to limitations of the basis and a severe truncation of the virtual space. The difference between diatomic and hydride $\gamma_M$ values in Table 2 suggests the present results are near the limits in accuracy of the method. Averaging over a different set of reference molecules to find better average values for $\gamma_M$ can improve individual molecules but is unlikely to produce significant improvements if applied to the entire set. It may be possible to improve the consistency by differentiating between *ss* and *pp* populations, but this has not been investigated.

**Table 4.** Comparison of calculated molecular total energies with exact values. Energies are reported for SCF and CI calculations with no correction included and for calculations including $\gamma$ in both the SCF and CI. Values for $\gamma$ are from Table 2. If the molecule does not contain hydrogen the diatomic value for $\gamma$ is used.

| | Total E Exact[a] | Atomization energy | SCF | CI[b] | Error CI - Exact | SCF | CI[b] | Error CI - Exact |
|---|---|---|---|---|---|---|---|---|
| | | | _______ no correction _______ | | | ____ includes correction[c] ______ | | |
| C | -37.85577 | | | | | | | |
| N | -54.61160 | | | | | | | |
| O | -75.10980 | | | | | | | |
| F | -99.80707 | | | | | | | |
| H | -0.50000 | | | | | | | |
| $H_2$ | -1.17447 | 0.17447 | -1.1311 | -1.1537 | 0.0207 | -1.1518 | -1.1745 | 0.0000 |
| $C_2$ | -75.94423 | 0.23269 | -75.5120 | -75.6397 | 0.3046 | -75.8166 | -75.9442 | 0.0000 |
| $CH_4$ | -40.52437 | 0.66860 | -40.2088 | -40.3226 | 0.2018 | -40.4008 | -40.5132 | 0.0112 |
| $N_2$ | -109.58714 | 0.36394 | -108.9885 | -109.2044 | 0.3827 | -109.3712 | -109.5871 | 0.0000 |
| $NH_3$ | -56.58549 | 0.47389 | -56.2130 | -56.3359 | 0.2496 | -56.4502 | -56.5715 | 0.0140 |
| $O_2$ | -150.41118 | 0.19159 | -149.6477 | -149.9012 | 0.5100 | -150.1577 | -150.4111 | 0.0001 |
| $H_2O$ | -76.47989 | 0.37009 | -76.0533 | -76.1842 | 0.2957 | -76.3417 | -76.4713 | 0.0086 |
| $F_2$ | -199.67501 | 0.06088 | -198.7537 | -199.0422 | 0.6328 | -199.3865 | -199.6750 | 0.0000 |
| HF | -100.53190 | 0.22483 | -100.0605 | -100.1970 | 0.3349 | -100.3947 | -100.5304 | 0.0015 |
| $C_2H_4$ | -78.60833 | 0.89679 | -78.0582 | -78.2581 | 0.3503 | -78.4167 | -78.6148 | -0.0065 |
| $C_2H_2$ | -77.35683 | 0.64529 | -76.8411 | -77.0394 | 0.3174 | -77.1836 | -77.3801 | -0.0232 |
| $C_2H_6$ | -79.84529 | 1.13375 | -79.2533 | -79.4598 | 0.3855 | -79.6231 | -79.8277 | 0.0176 |
| CO | -113.37868 | 0.41311 | -112.7807 | -112.9866 | 0.3921 | -113.1887 | -113.3938 | -0.0151 |
| $H_2CO$ | -114.56070 | 0.59513 | -113.9009 | -114.1242 | 0.4365 | -114.3454 | -114.5670 | -0.0063 |
| HCN | -93.46499 | 0.49763 | -92.9094 | -93.1145 | 0.3505 | -93.2864 | -93.4906 | -0.0256 |
| NO | -129.96452 | 0.24312 | -129.2909 | -129.5128 | 0.4517 | -129.7373 | -129.9588 | 0.0057 |
| $C_6H_6$ | -232.31284 | 2.17822 | -230.7684 | -231.2930 | 1.0199 | -231.7965 | -232.3193 | -0.0065 |
| $C_4H_4N_2$ | -264.42425 | 1.77798 | -262.7627 | -263.3116 | 1.1126 | -263.8573 | -264.4040 | 0.0202 |
| $C_5H_5N$ | -248.37746 | 1.98701 | -246.7701 | -247.3001 | 1.0773 | -247.8296 | -248.3571 | 0.0203 |
| $NH_2CH_2COOH$ | -284.59469 | 1.55195 | -282.9319 | -283.4429 | 1.1518 | -284.0439 | -284.5513 | 0.0434 |
| $C_6H_5NH_2$ | -287.71772 | 2.47150 | -285.8160 | -286.4294 | 1.2883 | -287.0638 | -287.6716 | 0.0461 |
| FHCO | -213.91366 | 0.64102 | -212.8353 | -213.1637 | 0.7499 | -213.5914 | -213.9181 | -0.0044 |
| $CF_2CH_2$ | -277.25192 | 0.92625 | -275.8541 | -276.2696 | 0.9824 | -276.8356 | -277.2494 | 0.0026 |
| $C_6H_5F$ | -331.65050 | 2.20881 | -329.6651 | -330.2875 | 1.3630 | -331.0091 | -331.6329 | 0.0176 |
| CHOCHO | -227.93995 | 1.00881 | -226.6750 | -227.0830 | 0.8570 | -227.5470 | -227.9524 | -0.0124 |
| $CH_2CHCHCH_2$ | -156.03489 | 1.61181 | -154.9703 | -155.3405 | 0.6944 | -155.6671 | -156.0346 | 0.0002 |
| $CH_3CH_2OH$ | -155.11080 | 1.28946 | -154.1331 | -154.4492 | 0.6616 | -154.7734 | -155.0868 | 0.0240 |
| $C_4H_4O$ | -230.11576 | 1.58288 | -228.6934 | -229.1623 | 0.9535 | -229.6390 | -230.1054 | 0.0104 |
| C₆H₅OH | -307.60522 | 2.36080 | -305.6569 | -306.2768 | 1.3284 | -306.9463 | -307.5690 | 0.0362 |
| HOOH | -151.64696 | 0.42736 | -150.8316 | -151.0820 | 0.5650 | -151.3842 | -151.6330 | 0.0139 |
| HNNH | -110.69009 | 0.46690 | -110.0378 | -110.2630 | 0.4271 | -110.4709 | -110.6947 | -0.0047 |
| N₂H₄ | -111.92037 | 0.69718 | -111.2090 | -111.4425 | 0.4779 | -111.6638 | -111.8950 | 0.0254 |
| HNO | -130.54823 | 0.32683 | -129.8372 | -130.0737 | 0.4745 | -130.3177 | -130.5531 | -0.0049 |
| HONO | -205.82788 | 0.49668 | -204.7183 | -205.0790 | 0.7489 | -205.4636 | -205.8229 | 0.0050 |
| CO₂ | -188.69544 | 0.62007 | -187.7097 | -188.0277 | 0.6678 | -188.3757 | -188.6921 | 0.0033 |
| CF₂ | -237.87679 | 0.40689 | -236.7607 | -237.0855 | 0.7913 | -237.5476 | -237.8709 | 0.0059 |
| CH₃F | -139.84221 | 0.67937 | -139.0966 | -139.3213 | 0.5209 | -139.6027 | -139.8259 | 0.0164 |
| HOF | -175.67171 | 0.25484 | -174.8098 | -175.0672 | 0.6045 | -175.4046 | -175.6611 | 0.0106 |
| CHF₃ | -338.50601 | 0.72904 | -336.9161 | -337.3613 | 1.1447 | -338.0464 | -338.4898 | 0.0162 |
| OF₂ | -274.87167 | 0.14774 | -273.5650 | -273.9722 | 0.8995 | -274.4534 | -274.8604 | 0.0113 |
| NO₂ | -205.19294 | 0.36174 | -204.1006 | -204.4678 | 0.7251 | -204.8033 | -205.1699 | 0.0230 |

[a] All energies in the table are in hartree units.
[b] The virtual space is truncated by omitting higher spherical harmonic functions (see text).
[c] Energy defect factors are from Table 2.

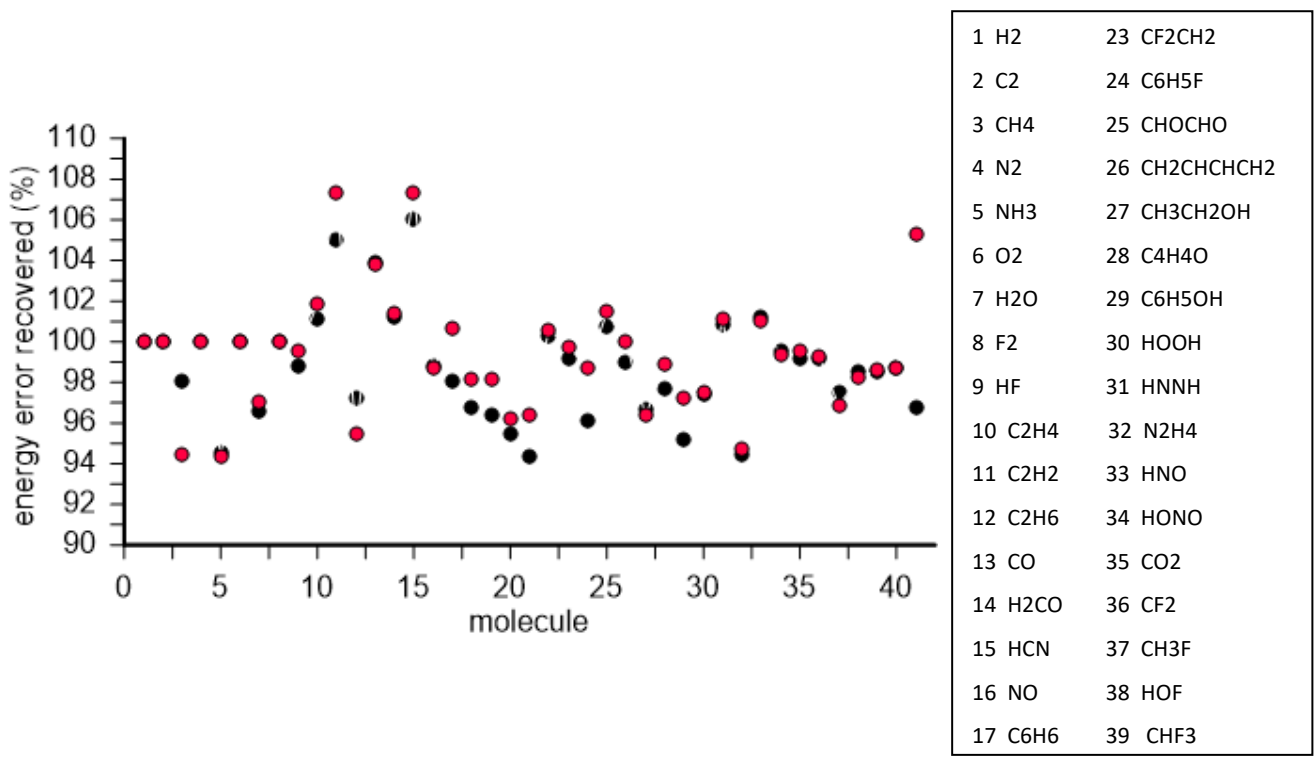

**Figure 1.** Energy error recovered by including γ in SCF and CI calculations (red) and by summing over invariant atomic error components Δ (black) from Tables 4 and 5, respectively.

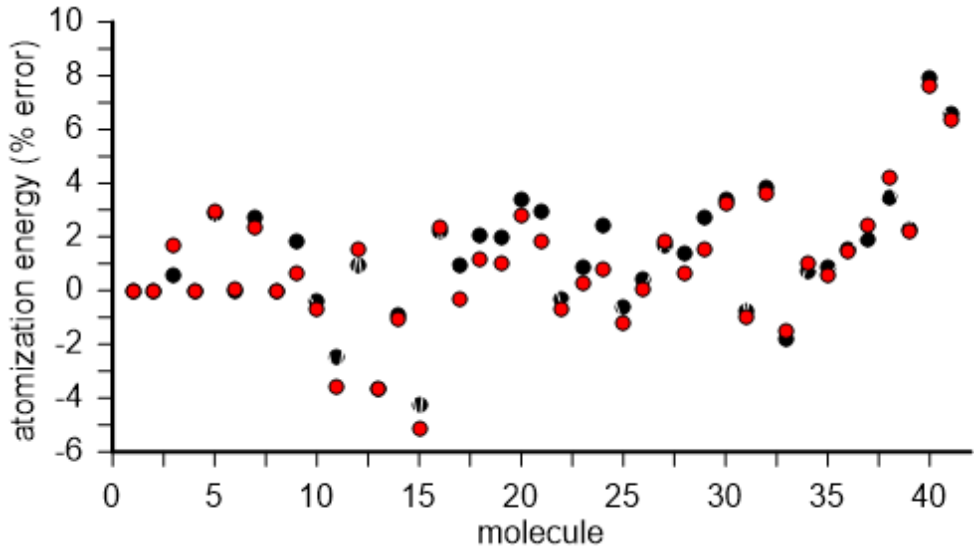

| 1 H2 | 23 CF2CH2 |
|---|---|
| 2 C2 | 24 C6H5F |
| 3 CH4 | 25 CHOCHO |
| 4 N2 | 26 CH2CHCHCH2 |
| 5 NH3 | 27 CH3CH2OH |
| 6 O2 | 28 C4H4O |
| 7 H2O | 29 C6H5OH |
| 8 F2 | 30 HOOH |
| 9 HF | 31 HNNH |
| 10 C2H4 | 32 N2H4 |
| 11 C2H2 | 33 HNO |
| 12 C2H6 | 34 HONO |
| 13 CO | 35 CO2 |
| 14 H2CO | 36 CF2 |
| 15 HCN | 37 CH3F |
| 16 NO | 38 HOF |
| 17 C6H6 | 39 CHF3 |

**Figure 2.** Atomization energy error by including γ in SCF and CI calculations (red) and by summing over invariant atomic error components Δ (black); from Tables 4 and 5, respectively. Atomization is to exact atoms thus there is no cancellation of errors in the calculation.

We now consider a simplification. Since partitioning the defect contribution of an overlap density is equivalent to using atomic populations and atomic defect factors, $\gamma_M$, it follows that the total correction for diatomic and hydride molecules can be restated exactly using neutral atom energy defect factors $\gamma'_M$.

For $M_2$, $\gamma = \gamma'$
and for $MH_y$ with $H$ population $1+\lambda$

$$E^{corr} = \gamma_M(n_M - y\lambda) + y(1+\lambda)\gamma_H = \gamma'_M n_M + y\gamma_H$$

where $n_M$ is the number of electrons of neutral $M$.

$$\gamma'_M = \gamma_M + y\lambda(\gamma_H - \gamma_M)/n_M$$

Thus, $\gamma'_M$ for the hydride is determined by $\gamma_M$, $\gamma_H$ and the charge transfer. The correction for a molecule containing $N_M$ atoms of type $M$ becomes

$$E^{corr} = \sum_M N_M n_M \gamma'_M = \sum_M N_M \Delta'_M \qquad (\Delta'_M = n_M \gamma'_M)$$

We refer to this simplification as the invariant atom approximation in which an average population is incorporated into the defect factor. Differences in the correction between the invariant atom and $h''$ methods are due primarily to the effect of $h''$ on the SCF molecular orbitals and energy.

Calculations using the invariant atom defect factors are reported in Table 5 and in Figures 1 and 2. The tables and figures show that correction energies and atomization energy errors are comparable for the two methods. The invariant atom method is slightly better for small molecules and the $h''$ method is slightly better for the larger systems where the SCF change is more important. Given the small size of the initial CI, it is encouraging that both methods work well.

In general, as an electronic structure calculation improves toward completeness the error correction decreases as will the differences between the diatomic and hydride $\gamma$ values, and improved consistency of the correction is expected. In the next section, we consider examples of coupled cluster calculations where the differences between the diatomic and hydride corrections are smaller, and the accuracy of the defect corrections is found to be improved.

**Table 5.** Energy defect correction based on invariant atomic contributions $\Delta'_M$. Energy defect factors are not included in SCF or CI calculations and the correction is $E^{corr} = \sum_M N_M \Delta'_M$ where $N_M$ is the number of atoms of type $M$. If the molecule does not contain hydrogen, the diatomic value is used for $\Delta'_M$.

|   | diatomic $\Delta'_M$ | $\Delta'_M$ | $\gamma'_M$ [a] |
|---|---|---|---|
| C | 0.15228 | 0.15630 | 0.02605 |
| N | 0.19137 | 0.20494 | 0.02928 |
| O | 0.25499 | 0.26496 | 0.03312 |
| F | 0.31640 | 0.32048 | 0.03561 |
| H | 0.01037 | 0.01037 | 0.01037 |

|   | Total E exact | Atomization exact | SCF | CI[b] | Error (CI-exact) | correction $E_{corr}$ | CI - exact (corrected) |
|---|---|---|---|---|---|---|---|
| H$_2$ | -1.17447 | 0.17447 | -1.1311 | -1.1537 | 0.0207 | 0.0207 | 0.0000 |
| C$_2$ | -75.94423 | 0.23269 | -75.5120 | -75.6397 | 0.3046 | 0.3046 | 0.0000 |
| CH$_4$ | -40.52437 | 0.66860 | -40.2088 | -40.3226 | 0.2018 | 0.1978 | 0.0040 |
| N$_2$ | -109.58714 | 0.36394 | -108.9885 | -109.2044 | 0.3827 | 0.3827 | 0.0000 |
| NH$_3$ | -56.58549 | 0.47389 | -56.2130 | -56.3359 | 0.2496 | 0.2361 | 0.0136 |

(middle header row: ____calculated (no correction)____)

| Molecule | | | | | | | |
|---|---|---|---|---|---|---|---|
| O$_2$ | -150.41118 | 0.19159 | -149.6477 | -149.9012 | 0.5100 | 0.5100 | 0.0000 |
| H$_2$O | -76.47989 | 0.37009 | -76.0533 | -76.1842 | 0.2957 | 0.2857 | 0.0100 |
| F$_2$ | -199.67501 | 0.06088 | -198.7537 | -199.0422 | 0.6328 | 0.6328 | 0.0000 |
| HF | -100.53190 | 0.22483 | -100.0605 | -100.1970 | 0.3349 | 0.3308 | 0.0041 |
| C$_2$H$_4$ | -78.60833 | 0.89679 | -78.0582 | -78.2581 | 0.3503 | 0.3541 | -0.0038 |
| C$_2$H$_2$ | -77.35683 | 0.64529 | -76.8411 | -77.0394 | 0.3174 | 0.3333 | -0.0159 |
| C$_2$H$_6$ | -79.84529 | 1.13375 | -79.2533 | -79.4598 | 0.3855 | 0.3748 | 0.0106 |
| CO | -113.37868 | 0.41311 | -112.7807 | -112.9866 | 0.3921 | 0.4073 | -0.0152 |
| H$_2$CO | -114.56070 | 0.59513 | -113.9009 | -114.1242 | 0.4365 | 0.4420 | -0.0055 |
| HCN | -93.46499 | 0.49763 | -92.9094 | -93.1145 | 0.3505 | 0.3716 | -0.0211 |
| NO | -129.96452 | 0.24312 | -129.2909 | -129.5128 | 0.4517 | 0.4464 | 0.0054 |
| C$_6$H$_6$ | -232.31284 | 2.17822 | -230.7684 | -231.2930 | 1.0199 | 1.0000 | 0.0198 |
| C$_4$H$_4$N$_2$ | -264.42425 | *1.77798* | -262.7627 | -263.3116 | 1.1126 | 1.0766 | 0.0361 |
| C$_5$H$_5$N | -248.37746 | *1.98701* | -246.7701 | -247.3001 | 1.0773 | 1.0383 | 0.0391 |
| NH$_2$CH$_2$COOH | -284.59469 | *1.55195* | -282.9319 | -283.4429 | 1.1518 | 1.0993 | 0.0525 |
| C$_6$H$_5$NH$_2$ | -287.71772 | *2.47150* | -285.8160 | -286.4294 | 1.2883 | 1.2153 | 0.0730 |
| FHCO | -213.91366 | *0.64102* | -212.8353 | -213.1637 | 0.7499 | 0.7521 | -0.0022 |
| CF$_2$CH$_2$ | -277.25192 | 0.92625 | -275.8541 | -276.2696 | 0.9824 | 0.9743 | 0.0081 |
| C$_6$H$_5$F | -331.65050 | 2.20881 | -329.6651 | -330.2875 | 1.3630 | 1.3101 | 0.0529 |
| CHOCHO | -227.93995 | 1.00881 | -226.6750 | -227.0830 | 0.8570 | 0.8633 | -0.0063 |
| CH$_2$CHCHCH$_2$ | -156.03489 | 1.61181 | -154.9703 | -155.3405 | 0.6944 | 0.6874 | 0.0070 |
| CH$_3$CH$_2$OH | -155.11080 | 1.28946 | -154.1331 | -154.4492 | 0.6616 | 0.6398 | 0.0218 |
| C$_4$H$_4$O | -230.11576 | 1.58288 | -228.6934 | -229.1623 | 0.9535 | 0.9316 | 0.0218 |
| C$_6$H$_5$OH | -307.60522 | 2.36080 | -305.6569 | -306.2768 | 1.3284 | 1.2650 | 0.0634 |
| HOOH | -151.64696 | 0.42736 | -150.8316 | -151.0820 | 0.5650 | 0.5507 | 0.0143 |
| HNNH | -110.69009 | 0.46690 | -110.0378 | -110.2630 | 0.4271 | 0.4306 | -0.0036 |
| N$_2$H$_4$ | -111.92037 | 0.69718 | -111.2090 | -111.4425 | 0.4779 | 0.4514 | 0.0265 |
| HNO | -130.54823 | 0.32683 | -129.8372 | -130.0737 | 0.4745 | 0.4803 | -0.0058 |
| HONO | -205.82788 | 0.49668 | -204.7183 | -205.0790 | 0.7489 | 0.7452 | 0.0037 |
| CO$_2$ | -188.69544 | 0.62007 | -187.7097 | -188.0277 | 0.6678 | 0.6623 | 0.0055 |
| CF$_2$ | -237.87679 | 0.40689 | -236.7607 | -237.0855 | 0.7913 | 0.7851 | 0.0062 |
| CH$_3$F | -139.84221 | 0.67937 | -139.0966 | -139.3213 | 0.5209 | 0.5079 | 0.0130 |
| HOF | -175.67171 | 0.25484 | -174.8098 | -175.0672 | 0.6045 | 0.5958 | 0.0087 |
| CHF$_3$ | -338.50601 | 0.72904 | -336.9161 | -337.3613 | 1.1447 | 1.1281 | 0.0167 |
| OF$_2$ | -274.871667 | 0.14774 | -273.5650 | -273.9722 | 0.8995 | 0.8878 | 0.0117 |
| NO$_2$ | -205.19294 | 0.36174 | -204.1006 | -204.4678 | 0.7251 | 0.7014 | 0.0237 |

[a] Included for comparison with values in Table 2. Only $\Delta'_M$ and the diatomic value are used for calculations.

[b] Higher spherical harmonic functions omitted from virtual space.

**V. Application to NIST database of coupled-cluster calculations**

In this section, we apply the invariant atom energy defect correction to a NIST database of coupled cluster CCSD(T)-full calculations carried out using a standard cc-pVTZ basis[2]. No new SCF or CI calculations are performed, and the analysis uses only the reported total energies. The energy correction is determined completely by the diatomic and hydride total energies. We have selected all molecules considered earlier for which the cc-pVTZ CCSD(T)-full calculations are reported in the database.

The objective is the same as in the truncated CI studies discussed previously using the invariant atom approximation: to correct the defect in the calculated CI energy to match the exact. For a given molecule, the correction is the same as in the previous section,

$$E^{corr} = \sum_M N_M n_M \gamma'_M = \sum_M N_M \Delta'_M \qquad (\Delta'_M = n_M \gamma'_M)$$

where $N_M$ is the number of atoms of type $M$ and $\Delta'_M$ is the average of the diatomic and hydride values. The key question is whether these two systems are sufficient to determine corrections accurately. In Table 6, energies before and after correction are compared with exact energies for all molecules investigated. Several points are noteworthy. Since the basis sets and virtual spaces are larger, the calculated coupled cluster energies are lower than from the truncated virtual space CI calculations reported in Tables 4 and 5. It follows that the $\Delta'_M$ values must be smaller than those reported in Table 5, More precisely, $\Delta'_M$ values in Table 6 are a factor of ~2 smaller for C,N, O, F and much smaller for H. However, the coupled cluster energies for the cc-pVTZ basis still show significant differences compared to exact energies. Including the energy defect correction greatly reduces the error. The table shows corrected energies in exceptionally good agreement with exact values for all molecules except acetylene. For this molecule, the tabulated cc-pVTZ value is inconsistent with the larger basis cc-pVQZ result also shown in Table 5. Using the latter value brings the corrected energy into good agreement with the exact value. In general, as a CI treatment improves toward completeness the error correction will decrease, as will the differences between the diatomic and hydride values of $\Delta'_M$, and one should expect increased reliability of the correction. The coupled cluster results which show excellent consistency support this conclusion.

**Table 6.** Analysis of NIST data base of coupled cluster CI calculations.[a] The energy defect correction for invariant atomic contributions $E^{corr} = \sum_M N_M n_M \gamma'_M = \sum_M N_M \Delta'_M$     $(\Delta'_M = n_M \gamma'_M)$   where $N_M$ is the number of atoms of type $M$.

| | Exact Energy[b] | $E_{CC}$ cc-pVTZ CCSD(T)=full | Error $E_{CC}$-Exact | $\Delta'_M$ [c] | Correction $E^{corr}$ | Error $E_{CC}$-$E^{corr}$-Exact |
|---|---|---|---|---|---|---|
| H$_2$ | -1.17447 | -1.1723 | 0.0021 | 0.001068 | | |
| C$_2$ | -75.94423 | -75.8071 | 0.1371 | 0.068569 (0.066782) | 0.1336 | 0.0036 |
| CH$_4$ | -40.52437 | -40.4551 | 0.0693 | 0.064995 | 0.0711 | -0.0018 |
| N$_2$ | -109.58714 | -109.3999 | 0.1872 | 0.093620 (0.093806) | 0.1876 | -0.0004 |
| NH$_3$ | -56.58549 | -56.4883 | 0.0972 | 0.093991 | 0.0970 | 0.0002 |
| O$_2$ | -150.41118 | -150.1536 | 0.2576 | 0.128782 (0.130378) | 0.2608 | -0.0032 |
| H$_2$O | -76.47989 | -76.3458 | 0.1341 | 0.131974 | 0.1325 | 0.0016 |
| F$_2$ | -199.67501 | -199.3205 | 0.3545 | 0.177249 (0.178363) | 0.3567 | -0.0022 |
| HF | -100.53190 | -100.3514 | 0.1805 | 0.179477 | 0.1794 | 0.0011 |
| C$_2$H$_4$ | -78.60833 | -78.4707 | 0.1377 | | 0.1378 | -0.0002 |
| C$_2$H$_2$ | -77.35683 | -77.1451 | 0.2117 | | 0.1357 | 0.0760 |
| C$_2$H$_2$ cc-pVQZ[d] | -77.35683 | -77.2674 | 0.0894 | | 0.0870 | 0.0023 |
| C$_2$H$_6$ | -79.84529 | -79.7079 | 0.1374 | | 0.1400 | -0.0026 |
| CO | -113.37868 | -113.1805 | 0.1982 | | 0.1972 | 0.0010 |
| H$_2$CO | -114.56070 | -114.3625 | 0.1982 | | 0.1993 | -0.0011 |
| HCN | -93.46499 | -93.3036 | 0.1614 | | 0.1617 | -0.0002 |
| NO | -129.96452 | -129.7420 | 0.2225 | | 0.2242 | -0.0017 |
| C$_6$H$_6$ | -232.31284 | -231.9024 | 0.4104 | | 0.4071 | 0.0033 |
| CH$_2$CHCHCH$_2$ | -156.03489 | -155.7556 | 0.2793 | | 0.2735 | 0.0058 |
| HOOH | -151.64696 | -151.3845 | 0.2625 | | 0.2629 | -0.0004 |
| HNNH | -110.69009 | -110.5054 | 0.1846 | | 0.1897 | -0.0051 |
| N$_2$H$_4$ | -111.92037 | -111.7278 | 0.1925 | | 0.1919 | 0.0007 |
| HNO | -130.54823 | -130.3245 | 0.2237 | | 0.2253 | -0.0016 |
| HONO | -188.69544 | -188.3683 | 0.3271 | | 0.3275 | -0.0004 |
| CO$_2$ | -237.87679 | -237.4567 | 0.4201 | | 0.4235 | -0.0034 |
| CF$_2$ | -139.84221 | -139.5872 | 0.2550 | | 0.2484 | 0.0066 |

| | | | | | |
|---|---|---|---|---|---|
| CH₃F | -175.67171 | -175.3343 | 0.3374 | 0.3098 | 0.0276 |
| NO₂ | -205.19294 | -204.8398 | 0.3531 | 0.3546 | -0.0014 |

[a] NIST database Ref. 2.
[b] Energies in hartrees
[c] Values of $\Delta'_M$ that correct the calculated CI energy to give the exact energy are given for diatomic molecules M₂ and hydride M_y. $\Delta'_M$ = error/2 and $\Delta'_M$ =error -y $\Delta'_H$, respectively. Corrections are calculated using the average values in parentheses.
[d] The second value for HCCH is from a larger basis set cc-pVQZ CCSD(T)=full calculation; the value reported for the smaller basis appears inconsistent.

## VI. Transition metal atomic states

We conclude the present study with an application to the $s^2d^n$, $sd^{n+1}$ and $d^{n+2}$ states of the first-row transition metals Sc-Cu. These states differ in their spatial orbitals and electron correlation. The basis is reported in Appendix A. Table 7 shows the result of SCF and CI calculations of the atomic states. Transition energies calculated by CI with no truncation of the virtual space differ from experimental values by 0.1 - 0.6 eV depending on the atom and state. We now introduce energy defect contributions, $\gamma$. Since the principal errors involve correlation associated with the d shell, it is necessary to distinguish between *ss, pp* and *dd* contributions to the density. The simplest choice is found to be satisfactory: to determine a value only for $\gamma_d$ and set $\gamma_{s,p}=0$. This means that the energy defect contributions have no effect on 1s, 2p, 2s,3s,3p and 4s electrons except indirectly due to changes in the d-shell. Cu with ground state $d^{10}s$ is an exception, where $\gamma_s \neq 0$, $\gamma_d=0$, and the correction is applied only to 4s electrons. Introducing the energy defect correction factors via $h''$ in both the SCF and CI gives the corrected energies reported in Table 7. The $\gamma$ value determined for each atom is also included in the table. The table shows considerable improvement in transition energies for all states with corrected differences from experiment reduced to 0.00 - 0.03eV.

Although not pursued in the present work, the plan is to use the $\gamma$ factors without change to describe states of molecules involving these atoms.

**Table 7.** Transition metal atomic states. Calculated transition energies are compared with spectroscopic values. Initial SCF and CI energies are reported along with recalculated energies from the inclusion of the defect energy correction in both the SCF and CI.

| | Expt. transition energy (eV)[a] | Calculated (no correction) | | | Calculated (includes correction) | | |
|---|---|---|---|---|---|---|---|
| | | SCF | CI | Transition energy (eV) | SCF | CI | Transition energy (eV) |
| **Sc** | | | | | | $\gamma = 0.009$ | |
| $^2$D ds$^2$ | | -759.7308 | -759.9525 | | -759.7398 | -759.9622 | |
| $^4$F d$^2$s | 1.43 | -759.6909 | -759.8909 | 1.68 | -759.7085 | -759.9092 | 1.44 |
| $^4$F d$^3$ | 4.19 | -759.5657 | -759.7813 | 4.66 | -759.5927 | -759.8082 | 4.19 |
| | | | | | | | |
| Ti | | | | | | $\gamma = 0.0055$ | |
| $^3$F d$^2$s$^2$ | | -848.3992 | -848.6197 | | -848.4102 | -848.6309 | |
| $^5$F d$^3$s | 0.81 | -848.3784 | -848.5845 | 0.96 | -848.3949 | -848.6013 | 0.81 |
| $^5$D d$^4$ | 3.57 | -848.2437 | -848.4770 | 3.88 | -848.2657 | -848.4992 | 3.59 |
| | | | | | | | |
| V | | | | | | $\gamma = 0.0016$ | |
| $^4$F d$^3$s$^2$ | | -942.8716 | -943.1042 | | -942.8764 | -943.1092 | |
| $^6$D d$^4$s | 0.26 | -942.8663 | -943.0933 | 0.30 | -942.8727 | -943.0997 | 0.26 |
| $^6$S d$^5$ | 2.51 | -942.7510 | -943.0080 | 2.62 | -942.7590 | -943.0161 | 2.53 |
| | | | | | | | |
| Cr | | | | | | $\gamma = 0.0080$ | |
| $^7$S d$^4$s$^2$ | 0.96 | -1043.3001 | -1043.5450 | 0.76 | -1043.3321 | -1043.5772 | 0.97 |
| $^5$D d$^5$s | 0.0 | -1043.3443 | -1043.5728 | 0.0 | -1043.3843 | -1043.6129 | 0.00 |
| $^5$D d$^6$ | 4.39 | -1043.0974 | -1043.4026 | 4.63 | -1043.1454 | -1043.4506 | 4.42 |
| | | | | | | | |
| Mn | | | | | | $\gamma = 0.0020$ | |
| $^6$S d$^5$s$^2$ | | -1149.8572 | -1150.0981 | | -1149.8672 | -1150.1077 | |
| $^6$D d$^6$s | 2.11 | -1149.7367 | -1150.0186 | 2.16 | -1149.7488 | -1150.0295 | 2.13 |
| $^4$P d$^7$ | 6.41 | -1149.5227 | -1149.8575 | 6.55 | -1149.5367 | -1149.8728 | 6.39 |
| | | | | | | | |
| Fe | | | | | | $\gamma = 0.0120$ | |
| $^5$D d$^6$s$^2$ | | -1262.4359 | -1262.7178 | | -1262.5079 | -1262.7899 | |
| $^5$F d$^7$s | 0.86 | -1262.3671 | -1262.6738 | 1.20 | -1262.4511 | -1262.7578 | 0.87 |
| ³F d⁸ | 4.08 | -1262.1622 | -1262.5443 | 4.72 | -1262.2582 | -1262.6403 | 4.07 |
| Co | | | | | $\gamma = 0.0120$ | | |
| ⁴F d⁷s² | | -1381.4024 | -1381.7143 | | -1381.4861 | -1381.7980 | |
| ⁴F d⁸s | 0.43 | -1381.3410 | -1381.6869 | 0.75 | -1381.4366 | -1381.7824 | 0.43 |
| ²D d⁹ | 3.41 | -1381.1388 | -1381.5643 | 4.08 | -1381.2463 | -1381.6718 | 3.43 |
| Ni | | | | | $\gamma = 0.0017$ | | |
| ³F d⁸s² | | -1506.8473 | -1507.1861 | | -1506.8605 | -1507.1993 | |
| ³D d⁹s | 0.025 | -1506.8038 | -1507.1836 | 0.07 | -1506.8184 | -1507.1984 | 0.025 |
| ¹S d¹⁰ | 1.83 | -1506.6484 | -1507.1150 | 1.94 | -1506.6649 | -1507.1314 | 1.822 |
| Cu | | | | | $\gamma_s = 0.0045$ | | |
| ²S d¹⁰s | | -1638.9402 | -1639.3643 | | -1639.0257 | -1639.4500 | |
| ²D d⁹s² | 1.39 | -1638.9236 | -1639.3083 | 1.52 | -1639.0226 | -1639.3985 | 1.40 |

**VII. Conclusions**

Methods for correcting residual energy errors of configuration interaction (CI) calculations of molecules and other electronic systems are discussed based on the assumption that the energy defect can be mapped onto atomic regions. It is assumed that the initial CI treatment adequately accounts for important non-local correlation contributions. Corrections are based on the premise that missing correlation and basis set contributions are of the same type as occur in smaller systems and can be recovered by understanding energy defects of the smaller systems.

It is shown that corrections determined by calculations only on diatomic and hydride molecules are sufficient to enable the correction of CI energies of larger molecules. This conclusion is supported by CI calculations on a test set of 41 molecules using two methods (inclusion of $h''$ and the invariant atom simplification) in the limit of a severely truncated virtual space. Both correction methods recover an average of ~98% of the initial energy defect and bring calculated CI energies into close agreement with exact thermodynamic energies. The simplified method is also applied to a NIST compilation of cc-pVTZ CCSD(T)-full coupled calculations that employ a larger basis set and no truncation of the virtual space. The corrections show excellent consistency and total energies are in very good agreement with experimental values. An extension of the method is applied to $d^m s^n$ states of Sc, Ti, V, Mn, Cr, Fe, Co, Ni and Cu, significantly improving the agreement of transition energies with spectroscopic values

The present results are encouraging and suggest that it would be useful to obtain additional data from new electronic structure calculations by routinely adding the invariant atom correction which requires only diatomic and hydride energies. Alternatively, instead of targeting the exact energy, corrections could be determined by the same procedure to estimate the energy of a higher-level CI treatment starting with a lower-level calculation.

## VIII. Appendix

### Basis set

The basis for each atom is a near Hartree-Fock set of atomic orbitals plus extra two-component s- and p-type functions consisting of the two smaller exponent components of the atomic orbital; sets of two-component d and two-component p functions are added for first-row atoms and hydrogen, respectively. The latter d- and p-type functions were optimized by CI calculations on atoms. Orbitals are expanded as linear combinations of Gaussian functions: 1s(10), 2s(5), 2p(6), s′(2), s″(1), p′(2), p″(1), d(2), for C,N,O , 2p(7) for F, and 1s(4), s(1), p(2) for H where the number of Gaussian functions in each orbital is indicated in parentheses. The transition metal basis is 1s(12), 2s(10), 2p(7), 3s(7), 3p(6), 4s(4), 3d(5), d′(4), d″(2), s′(1), s″(1), p′(2), p″(2). No core potentials are used in the present calculations.

### Configuration interaction

All calculations are carried out for the full electrostatic Hamiltonian of the system

$$H = \sum_{i}^{N}[-\tfrac{1}{2}\nabla_i^2 + \sum_{k}^{Q} -\frac{Z_k}{r_{ik}}] + \sum_{i<j}^{N} r_{ij}^{-1}$$

A single-determinant self-consistent-field (SCF) solution is obtained initially for each state of interest. Configuration interaction wavefunctions are constructed by multi-reference expansions,[7-8]

$$\Psi = \sum_{k} c_k (N!)^{-1/2} det(\chi_1^k \, \chi_2^k \ldots \chi_N^k) = \sum_{k} c_k \Phi_k$$

In all applications, the entire set of SCF orbitals is used to define the CI active space. Virtual orbitals are determined by a positive ion transformation to improve convergence. Single and double excitations from the single determinant SCF wavefunction, $\Phi_r$, creates a small CI expansion, $\Psi'_r$,

$$\Psi'_r = \Phi_r + \sum_{ijkl} \lambda_{ijkl} \Gamma_{ij \to kl} \Phi_r = \sum_m c_m \Phi_m$$

The configurations $\Phi_m$, are retained if the interaction with $\Phi_r$ satisfies a relatively large second order energy condition

$$\frac{|<\Phi_m|H|\Phi_r>|^2}{E_m - E_r + \lambda} \geq 10^{-4}$$

The description is then refined by generating a large CI expansion, $\Psi_r$ by single and double excitations from all important members of $\Psi'_r$ to obtain

$$\Psi_r = \Psi'_r + \sum_m \left[ \sum_{ik} \lambda_{ikm} \Gamma_{i \to k} \Phi_m + \sum_{ijkl} \lambda_{iklm} \Gamma_{ij \to kl} \Phi_m \right]$$

where $\Phi_m$ is a member of $\Psi'_r$ with coefficient $> 0.01$. Typically, $\Psi'_r$ contains 200-400 dets. We refer to this expansion as a multi-reference CI. The additional configurations are generated by identifying and retaining all configurations, $\Phi_m$, that interact with $\Psi'_r$ such that

$$\frac{|<\Phi_m|H|\Psi'_r>|^2}{E_m - E_r + \lambda} \geq 10^{-6}$$

For the molecules investigated, approximately $10^5$-$10^6$ determinants occur in the final CI expansion, and the expansion can contain single through quadruple excitations from an initial representation of the state $\Phi_r$. The contribution of determinants not explicitly included along with size consistency corrections are estimated by perturbation theory. The value of $\lambda$ is determined so that the second order perturbation energy matches the CI value if first order coefficients

$$c_m = \frac{-<\Phi_m|H|\Psi'_r>}{E_m - E_r + \lambda}$$

are used for determinants in the CI calculation.

**Experimental thermodynamic data from NIST compilation** [a]

|  | Atomizaton energy vibrational kJ | ZPE cm-1 | integrated Cp kJ | Exact energy hartree | Exact atomization incl ZPE hartree |
|---|---|---|---|---|---|
|  |  |  | 6.197 |  |  |
| C |  |  | 6.536 | -37.85577 |  |
| N |  |  | 6.197 | -54.61160 |  |
| O |  |  | 6.725 | -75.10980 |  |
| F |  |  | 6.518 | -99.80707 |  |
| H |  |  | 6.197 | -0.50000 |  |
| H$_2$ | 432.1 | 2179.3 | 8.468 | -1.17447 | 0.17447 |
| C$_2$ | 600 | 914 | 10.169 | -75.94423 | 0.23269 |
| CH$_4$ | 1642 | 9480 | 10.016 | -40.52437 | 0.66860 |
| N$_2$ | 941.6 | 1165 | 8.670 | -109.58714 | 0.36394 |
| NH$_3$ | 1157.9 | 7214.5 | 10.043 | -56.58549 | 0.47389 |
| O$_2$ | 493.7 | 778 | 8.680 | -150.41118 | 0.19159 |
| H$_2$O | 917.8 | 4504 | 9.905 | -76.47989 | 0.37009 |
| F$_2$ | 154.5 | 447 | 8.825 | -199.67501 | 0.06088 |
| HF | 566.6 | 1980.7 | 8.599 | -100.53190 | 0.22483 |
| C$_2$H$_4$ | 2225.5 | 10784.7 | 10.518 | -78.60833 | 0.89679 |
| C$_2$H$_2$ | 1626.5 | 5660.5 | 10.009 | -77.35683 | 0.64529 |
| C$_2$H$_6$ | 2787 | 15853.5 | 11.884 | -79.84529 | 1.13375 |
| CO | 1071.80 | 1071.6 | 8.671 | -113.37868 | 0.41311 |
| H$_2$CO | 1495 | 5643.5 | 10.020 | -114.56070 | 0.59513 |
| HCN | 1265.7 | 3412.5 | 9.235 | -93.46499 | 0.49763 |
| NO | 627.1 | 938 | 9.192 | -129.96452 | 0.24312 |
| C$_6$H$_6$ | 5463 | 21392.5 | 14.331 | -232.31284 | 2.17822 |
| C$_4$H$_4$N$_2$ | 4488 | 16307.5 | 15 [e] | -264.42425 | 1.77798 |
| C$_5$H$_5$N | 5005.7 | 18909.2 | 15 [e] | -248.37746 | 1.98701 |
| NH$_2$CH$_2$COOH | 3885 | 17107 [b] | 15 [e] | -284.59469 | 1.55195 |
| C$_6$H$_5$NH$_2$ | 6211.7 | 24527 [c] | 16.184 [e] | -287.71772 | 2.47150 |
| FHCO | 1639.8 | 4449.6 | 10.02 [e] | -213.91366 | 0.64102 |
| CF$_2$CH$_2$ | 2338.5 | 7805 | 12.048 | -277.25192 | 0.92625 |

| Molecule | | | | |
|---|---|---|---|---|
| $C_6H_5F$ | 5580.2 | 19663.5 | 16.184 | -331.65050 | 2.20881 |
| CHOCHO | 2554.5 | 7868 | 13.673 | -227.93995 | 1.00881 |
| $CH_2CHCHCH_2$ | 4016.5 | 17998.5 | 15.134 | -156.03489 | 1.61181 |
| $CH_3CH_2OH$ | 3182.50 | 16968.4 | 14.126 | -155.11080 | 1.28946 |
| $C_4H_4O$ | 3977.4 | 14918.5 | 12.347 | -230.11576 | 1.58288 |
| $C_6H_5OH$ | 5953.7 | 21798 [d] | 16.184 [e] | -307.60522 | 2.36080 |
| HOOH | 1055.5 | 5561.5 | 11.158 | -151.64696 | 0.42736 |
| HNNH | 1154.7 | 5947.7 | 9.997 | -110.69009 | 0.46690 |
| $N_2H_4$ | 1696.4 | 11204.9 | 11.449 | -111.92037 | 0.69718 |
| HNO | 823.7 | 2875 | 9.942 | -130.54823 | 0.32683 |
| HONO | 1253.3 | 4241.2 | 11.597 | -205.82788 | 0.49668 |
| $CO_2$ | 1598.00 | 2508 | 9.365 | -188.69544 | 0.62007 |
| $CF_2$ | 1050.3 | 1503.3 | 10.353 | -237.87679 | 0.40689 |
| $OF_2$ | 374.60 | 1110 | 10.895 | -274.87167 | 0.14774 |
| $CH_3F$ | 1683.5 | 8376 | 10.135 | -139.84221 | 0.67937 |
| HOF | 634.2 | 2916.8 | 10.088 | -175.67171 | 0.25484 |
| $CHF_3$ | 1848.8 | 5457.5 | 11.565 | -338.50601 | 0.72904 |
| $NO_2$ | 927.70 | 1843 | 10.186 | -205.19294 | 0.36174 |

[a] Ref. 2   [b] CCpVDZ scaled   [c] SDCI 6-31G* scaled   [d] CC 6-31G* scaled

[e] Estimated from molecules with similar structure.

**Data Availability**

Data used in this work are available on request.

**References**

1. Shavitt, Isaiah; Bartlett, Rodney J. (2009). Many-Body Methods in Chemistry and Physics: MBPT and Coupled-Cluster Theory. Cambridge University Press. ISBN 978-0-521-81832-2

**2.** NIST Computational Chemistry Comparison and Benchmark Database**,** NIST Standard Reference Database Number 101**,** Release 21, August 2020, Editor: Russell D. Johnson III http://cccbdb.nist.gov/ DOI:10.18434/T47C7

Kramida, A., Ralchenko, Yu., Reader, J., and NIST ASD Team (2021). *NIST Atomic Spectra Database* (ver. 5.9), [Online]. https://physics.nist.gov/asd [2022, May 5]. National Institute of Standards and Technology, Gaithersburg, MD. DOI: https://doi.org/10.18434/T4W30F

3. A Pipano and I. Shavitt, Intern. J. Quantum Chern. 2, 741 (1968); Z. Gershgorn and I. Shavitt, ibid. 2, 751 (1968).

4. Werner Kutzelnigg and Pasquale von Herigonte, Electron correlation at the dawn of the 21$^{st}$ century, in Advances in Quantum Chemistry, Volume 36, Pages 185-229 (2000).

5. Jerry L. Whitten, J. Chem. Phys. 153, 244103 (2020). doi: 10.1063/5.0031279.

6. Jiří Čížek, J. Chem. Phys., 45, 4256, (1966).

7. J. Paldus (2005). "The beginnings of coupled-cluster theory: an eyewitness account". In Dykstra, C. Theory and Applications of Computational Chemistry: The First Forty Years. Elsevier B.V. p. 115.

8. S. Bennie, M Van der Kamp, R Pennifold, M Stella, F. Manby and A. Mulholland, J. Chem. Theory Comp., 12, 2689 (2016).

9. J. L. Whitten and M. Hackmeyer, J. Chem. Phys. 51, 5584 (1969); B. N. Papas and J. L. Whitten, J. Chem. Phys., 135, 204701 (2011).

10. R. J. Buenker and S. D Peyerimhoff, Theor Chim Acta, 35, 33 (1974).

11. Stefano Evangelisti, Jean-Pierre Daudey, Jean-Paul Malrieu, Chemical Physics,

75, 91, 1983.

12. J. E Subotnik and M. Head-Gordon, J. Chem. Phys., 123, 064108 (2005); 122, 034109 (2005).

13. H. J. Werner and W. Meyer, *J. Chem. Phys.,* **73**, 2342 (1980).

14. P Pulay, S. Saebo, and W. Meyer, J. Chem. Phys., 81, 1901-5 (1984).

15. S. Saebo and P. Pulay, Chem. Phys. Lett. 113, 13 (1985). S. Saebo and P. Pulay, Fourth-Order MollerPlesset Perturbation Theory in the Local Correlation Treatment, I. Theory, J. Chem. Phys. 86, 914 (1987); II. Results, J. Chem. Phys., 88, 1884 (1988).

16. S. Saebo and P. Pulay, A low-scaling method for second order Møller-Plesset calculations, J. Chem. Phys., 115, 3975 (2001).

17. Filipp Furche, Reinhart Ahlrichs, Christof Hättig, Wim Klopper, Marek Sierka, and Florian Weigend, WIREs Computational Molecular Science, John Wiley & Sons, Ltd. (2013).

18. Peter Pulay, , Chemical Physics Letters, 100, 151 (1983).

19. R. A Friesner, R. B Murphy, M. D. Beachy, M. N. Ringnalda, W. T Pollard, B. D. Dunietz and Y. X. Cao, J. Phys. Chem. A, 103, 1913 (1999).

20. P. Y Ayala and G. E. Scuseria, J. Chem. Phys., 110, 3660 (1999)

21. For early work on electron correlation functionals for molecules see: G. C. Lie and E. Clementi, J. Chem. Phys., 60, 1288 (1974);  Enrico Clementi, Subhas J. Chakravorty, J. Chem. Phys., 93, 2591 (1990).

22. G. E Scuseria and P. Y., Ayala, J. Chem. Phys., 111, 8330 (1999).

23.  J.E. Deustua, I. Magoulas, J. Shen and P. Piecuch, J. Chem. Phys., *149,* 151101 **(2018)**; J.E. Deustua, S.H. Yuwono, J. Shen and P. Piecuch, J. Chem. Phys*., 150,* 111101 **(2019).**

24. Devin A. Matthews, Lan Cheng, Michael E. Harding, Filippo Lipparini, Stella Stopkowicz, Thomas-C., Szalay Jagau, G. Péter, Jürgen Gauss and John F Stanton, J. Chem. Phys., 152, 214108 (2020).

25. C. Song and T. J. Martinez 152, 234113 (2020); B. S. Fales and T. J.  Martinez, J. Chem. Phys., 152, 164111 (2020).

26. B. Scott Fales and Todd J. Martínez, J. Chem. Theory Comput., 16, 3, 1586–1596 (2020).

27. Thierry Leininger, Hermann Stoll, Hans-Joachim Werner and Andreas Savin, Chemical Physics Letters, 275, 151 (1997).

28. W. Meyer, J. Chem. Phys. 58, 1017 (1973).

29. R. K. Kathir, Coen de Graaf, Ria Broer, Remco W. A. Havenith, and J. Chem. Theory Comput., 16, 2941 (2020).

30. Jerry L Whitten, J. Chem. Phys. 146, 064113 (2017).

31. Jerry L Whitten, Phys. Chem. Chem. Phys., 21, 21541 (2019).

32. Clyde Edmiston and Klaus Ruedenberg, Reviews of Modern Physics, 35, 457 (1963).

33. J. M. Foster and S. F Boys, Reviews of Modern Physics, 32, 300 (1960).

34. J. L. Whitten and H. Yang Int. J. Quantum Chem.: Quantum Chemistry Symposium, 29, 41-47 (1995).

35. Susi Lehtola and Hannes Jónsson, J. Chem. Theory and Comput., 9, 5365–5372 (2013).

36. Ida-Marie Høyvik, Branislav Jansik and Poul Jørgensen, J. Chem. Theory and Comput., 8, 3137 (2012).

37. See for example: J. L. Whitten and H. Yang, Surface Science Reports, 24, 55-124 (1996) and Brian N. Papas and Jerry L. Whitten, Int. J. Quantum Chemistry, 110, 3072–3079 (2010) and references contained therein.

38. Ahmet Alturn, Robert Izsak and Givanni Bistoni, Int. J. Quantum Chemistry (2020) e26339.

39. Yang Guo, Wei Li and Shuhua Li, J. Phys. Chem. A, 118, 39, 8996 (2014).

40. P. Hohenberg and W. Kohn, Phys. Rev., 136, B864 (1964).

41. J. L. Whitten, J. Chem. Phys., 58, 4496 (1973).

42. J. A. Jafri and J. L. Whitten, J. Chem. Phys., 61, 2116 (1974); T.A. Pakkanen and J. L. Whitten, J. Chem. Phys. 69, 2168 (1978).

43. Hans-Joachim Werner and Martin Schütz, J. Chem. Phys., 135, 144116 (2011).

44. Frank Neese, Frank Wennmohs, Ute Becker, Christoph Riplingr, J. Chem. Phys., 152, 224108 (2020).

45. E. J. Baerends, D. E. Ellis, and P. Ros, Chem. Phys. **2**, 41 (1973).

46. B. I. Dunlap, J. W. D. Connolly, and J. R. Sabin, J. Chem. Phys. **71**, 3396 (1979).